\documentclass[%
reprint,
superscriptaddress,
preprintnumbers,
amsmath,amssymb,
prl,
aps,
floatfix
]{revtex4-1}

\usepackage{graphicx}
\usepackage{dcolumn}
\usepackage{bm}
\usepackage{hyperref}
\usepackage{comment}
\usepackage{placeins}

\usepackage[normalem]{ulem} 
\usepackage{xcolor}

\newcommand{\eh}{$e^-h^+$}
\newcommand{\eV}{$\mathrm{eV/c^2}$}
\newcommand{\keV}{$\mathrm{keV/c^2}$}
\newcommand{\MeV}{$\mathrm{MeV/c^2}$}
\newcommand{\GeV}{$\mathrm{GeV/c^2}$}
\newcommand{\gd}{g\,d}

\begin{document}

\title{First Dark Matter Constraints from a SuperCDMS Single-Charge Sensitive Detector}

\affiliation{Division of Physics, Mathematics, \& Astronomy, California Institute of Technology, Pasadena, CA 91125, USA}
\affiliation{Department of Physics, Durham University, Durham DH1 3LE, UK}
\affiliation{Fermi National Accelerator Laboratory, Batavia, IL 60510, USA}
\affiliation{Lawrence Berkeley National Laboratory, Berkeley, CA 94720, USA}
\affiliation{Department of Physics, Massachusetts Institute of Technology, Cambridge, MA 02139, USA}
\affiliation{School of Physical Sciences, National Institute of Science Education and Research, HBNI, Jatni - 752050, India}
\affiliation{Department of Physics \& Astronomy, Northwestern University, Evanston, IL 60208-3112, USA}
\affiliation{Pacific Northwest National Laboratory, Richland, WA 99352, USA}
\affiliation{Department of Physics, Queen's University, Kingston, ON K7L 3N6, Canada}
\affiliation{Department of Physics, Santa Clara University, Santa Clara, CA 95053, USA}
\affiliation{SLAC National Accelerator Laboratory/Kavli Institute for Particle Astrophysics and Cosmology, 2575 Sand Hill Road, Menlo Park 94025, CA}
\affiliation{Department of Physics, South Dakota School of Mines and Technology, Rapid City, SD 57701, USA}
\affiliation{Department of Physics, Southern Methodist University, Dallas, TX 75275, USA}
\affiliation{Department of Physics, Stanford University, Stanford, CA 94305, USA}
\affiliation{Department of Physics, Syracuse University, Syracuse, NY 13244, USA}
\affiliation{Department of Physics and Astronomy, and the Mitchell Institute for Fundamental Physics and Astronomy, Texas A\&M University, College Station, TX 77843, USA}
\affiliation{SNOLAB, Creighton Mine \#9, 1039 Regional Road 24, Sudbury, ON P3Y 1N2, Canada}
\affiliation{TRIUMF, Vancouver, BC V6T 2A3, Canada}
\affiliation{Instituto de F\'{\i}sica Te\'orica UAM/CSIC, Universidad Aut\'onoma de Madrid, 28049 Madrid, Spain}
\affiliation{D\'{e}partement de Physique, Universit\'{e} de Montr\'{e}al, Montr\'{e}al, QC  H3T 1J4, Canada}
\affiliation{Department of Physics \& Astronomy, University of British Columbia, Vancouver, BC V6T 1Z1, Canada}
\affiliation{Department of Physics, University of California, Berkeley, CA 94720, USA}
\affiliation{Department of Physics, University of California, Santa Barbara, CA 93106, USA}
\affiliation{Department of Physics, University of Colorado Denver, Denver, CO 80217, USA}
\affiliation{Department of Electrical Engineering, University of Colorado Denver, Denver, CO 80217, USA}
\affiliation{Department of Physics, University of Evansville, Evansville, IN 47722, USA}
\affiliation{Department of Physics, University of Florida, Gainesville, FL 32611, USA}
\affiliation{Department of Physics, University of Illinois at Urbana-Champaign, Urbana, IL 61801, USA}
\affiliation{School of Physics \& Astronomy, University of Minnesota, Minneapolis, MN 55455, USA}
\affiliation{Department of Physics, University of South Dakota, Vermillion, SD 57069, USA}
\affiliation{Department of Physics, University of Toronto, Toronto, ON M5S 1A7, Canada}

\author{R.~Agnese} \affiliation{Department of Physics, University of Florida, Gainesville, FL 32611, USA}
\author{T.~Aralis} \affiliation{Division of Physics, Mathematics, \& Astronomy, California Institute of Technology, Pasadena, CA 91125, USA}
\author{T.~Aramaki} \affiliation{SLAC National Accelerator Laboratory/Kavli Institute for Particle Astrophysics and Cosmology, 2575 Sand Hill Road, Menlo Park 94025, CA}
\author{I.J.~Arnquist} \affiliation{Pacific Northwest National Laboratory, Richland, WA 99352, USA}
\author{E.~Azadbakht} \affiliation{Department of Physics and Astronomy, and the Mitchell Institute for Fundamental Physics and Astronomy, Texas A\&M University, College Station, TX 77843, USA}
\author{W.~Baker} \affiliation{Department of Physics and Astronomy, and the Mitchell Institute for Fundamental Physics and Astronomy, Texas A\&M University, College Station, TX 77843, USA}
\author{S.~Banik} \affiliation{School of Physical Sciences, National Institute of Science Education and Research, HBNI, Jatni - 752050, India}
\author{D.~Barker} \affiliation{School of Physics \& Astronomy, University of Minnesota, Minneapolis, MN 55455, USA}
\author{D.A.~Bauer} \affiliation{Fermi National Accelerator Laboratory, Batavia, IL 60510, USA}
\author{T.~Binder} \affiliation{Department of Physics, University of South Dakota, Vermillion, SD 57069, USA}
% ...

\received{?}
\accepted{?}
\date{\today}

\setcounter{figure}{0}
\begin{abstract}
We present the first limits on inelastic electron-scattering dark matter and dark photon absorption using a prototype SuperCDMS detector having a charge resolution of 0.1 electron-hole pairs (CDMS HVeV, a 0.93 gram CDMS HV device). These electron-recoil limits significantly improve experimental constraints on dark matter particles with masses as low as 1~\MeV. We demonstrate a sensitivity to dark photons competitive with other leading approaches but using substantially less exposure (0.49 gram days). These results demonstrate the scientific potential of phonon-mediated semiconductor detectors that are sensitive to single electronic excitations. 
\end{abstract}

\maketitle

~\\

\section{Introduction}
Over the past few years, the LHC and direct detection experiments have ruled out a substantial portion of the most natural parameter space for Weakly Interacting Massive Particle (WIMP) dark matter (DM), motivating new searches for DM over a broader mass range (Ref.~\cite{CosmicVisions} and references therein). In particular, sub-\GeV~DM that couples to Standard Model (SM) particles through a new force mediator is a well motivated alternative to the WIMP hypothesis\cite{LightDarkSectors,DarkSectors,Essig2012}. \eV-scale bosonic dark matter in the form of dark photons \cite{Okun,Holdom,galiston} and \MeV-scale fermionic dark matter which forms the lightest particle in a new force sector \cite{Boehm:2003hm,Pospelov} are both capable of reproducing the dark matter relic density while evading current constraints \cite{An,Izaguirre,Essig}.

The unconstrained parameter space in these models can be probed by new gram-scale detectors with single charge resolution~\cite{Izaguirre,Essig,Hochberg,An}. Dark photon signatures may be probed through kinetic mixing with the SM photon and subsequent absorption by the detector\cite{Hochberg}. Low-mass DM interactions that excite electrons from bound to unbound states can efficiently transfer large fractions of the total DM kinetic energy to these electrons, making inelastic electron-recoil DM (ERDM) searches compelling~\cite{Essig}. The inelastic ERDM scattering rate depends strongly on both the size of the target material's band gap and the detection threshold. Consequently, a semiconductor detector with sensitivity to single electron-hole (\eh) pairs \cite{Romani, Tiffenberg_17prl_Sensei} can be competitive with other experimental technologies \cite{damic,Xenon10}, even with a very modest exposure in an above-ground facility.

In this paper we present results from our first sub-\GeV~ERDM and dark photon searches with 0.49 gram days (\gd) of exposure of the CDMS HVeV detector~\cite{Romani} (a gram-scale CDMS HV~\cite{Agnese_PRD17_SuperCDMSSensitivity} prototype with eV-scale resolution). We discuss the performance of this device, including the charge leakage measured during long exposures, and the path forward to future experiments with both silicon (Si) and germanium (Ge) detectors.


\section{Experimental Setup}

This search employed a $1{\times}1{\times}0.4$~cm$^3$ high-purity Si crystal (0.93~g) instrumented on one side with two channels of quasiparticle-trap-assisted electro-thermal-feedback transition-edge sensors (QETs), biased at -42\,mV, and on the other side with a 20\% coverage electrode consisting of an aluminum/amorphous silicon bilayer~\cite{Romani}, biased relative to ground. The QETs, which measure the total energy of phonons produced in the substrate, had an energy resolution of $\sigma_{ph} \sim 14$~eV at the nominal base temperature of 33--36~mK~\cite{Romani} (a significant advance for Si calorimetry comparable to that recently achieved in sapphire\cite{Strauss}). Single-charge resolution was achieved by drifting \eh\ pairs across 140\,V to amplify the small charge signal into a large phonon signal via the Neganov-Trofimov-Luke (NTL) effect~\cite{Neganov1985,Luke1988}. The bias voltage did not increase the baseline phonon noise, resulting in an effective charge resolution of $\sigma_{eh}=\frac{\sigma_{ph}}{qV} \approx$~0.1 \eh\ pairs, where $q$ is the quantum of charge and V the bias voltage. In Si, where the creation energy per \eh\ pair is $\epsilon_{eh}=3.8$~eV, this is equivalent to $\sim0.4$~eV in electronic recoil energy, though it is a discrete energy scale (see Eq.~\ref{eq:ionization}) and \eh\ pairs can be generated down to the band gap energy $E_{gap}=1.2$~eV~\cite{Vavilov}.

A pulsed monochromatic 650 nm laser ($\sim$1.91\,eV photons) provided periodic in-run calibrations, with a repetition rate of 1 Hz and an average number of photons absorbed per pulse, $\lambda$, of approximately 2. Data were acquired by triggering on a laser coincident logic signal for diagnostic studies or a shaped pulse---sum of the two QET channels through a shaping amplifier---for the science exposure. The trigger threshold for the shaped detector pulses was set to 0.5 \eh\ pairs based on pre-run calibration data. This resulted in a trigger efficiency $>$95\% for one \eh\ pair. Independent of the trigger source, the data recorded for each event consisted of the two unshaped QET responses, the laser coincidence signal, the shaped QET pulse, and the dilution refrigerator (DR) temperature.

In Ref. \cite{Romani}, we argued that sub-gap infrared photons (SGIR) excited neutralized impurities within the bulk of the crystal, producing unpaired excitations that would drift across only a fraction of the potential drop and thus have non-integer signal amplitude. For the data presented here, SGIR was mitigated with changes to the optical fiber coupled to the pulsed laser source and the introduction of IR filters, rated to reduce transmitted IR at 800~nm by $\sim$99.8\% and attenuate transmission at longer wavelengths by several orders of magnitude.

Data were acquired over 6 days with 36 hours of raw exposure comprising the selected data set: 27 hours at -140\,V and 9 hours at +140\,V. Only the negative-bias data were considered for this analysis, with some of the raw exposure removed because of sporadic periods of high electrical noise and drifts in base temperature above 36\,mK. Variations in the DR temperature led to gain variations resulting in a variable threshold between 0.2 and 0.5 \eh\ pairs. This temperature-correlated gain variation was corrected based on the known DR temperature, as described in the following section.

\section{Data Reconstruction and Calibration}
The full dataset was separated into series of 10,000 events. For each series, noise spectra and phonon pulse templates were computed from the first and second halves of the digitized traces respectively, with the trigger point located in the second half of the trace. The pulse amplitude and start time were estimated using the optimal filter formalism (e.g. Ref~\cite{Golwala}). We reconstructed amplitudes for individual channels and their sum in order to quantify signal position dependence and channel noise covariance. We observed time variation in the noise spectra and the pulse amplitude, but not in the shape of the templates. Thus a single averaged pulse template was generated from laser calibration events taken over the entire science exposure.

The laser calibration showed that the detector energy response was nonlinear, requiring a quadratic correction to convert from pulse height to an absolute energy scale as discussed in Ref.~\cite{Romani}. Additionally, the change in the overall energy scale caused by temperature variations was corrected by aligning the laser spectral peaks with equal \eh\ pair quanta. The temperature correction was observed to be linear in energy throughout our analysis region of 0--10 \eh\ pairs. Finally, we compared laser events with events from periods with elevated surface leakage near the outer edge of the detector to determine the relative calibration gain factor between the inner and outer QET channels. This resulted in a 30\% increase in the outer channel amplitude. The calibrated total energy is thus position and temperature independent. 

\begin{figure}[t]
\centering
\includegraphics[clip=True,trim=6 6 0 0,width=3.25in]{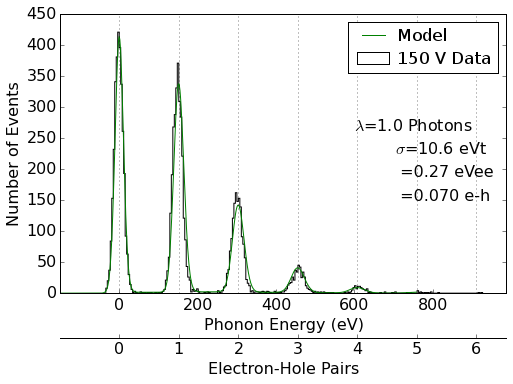}
\caption{Laser calibration data showing a resolution of ${\sim}$0.07~\eh pairs for a short laser-triggered acquisition at 150\,V. In the series shown in this figure, a lower DR temperature allowed for a 30\% improvement in energy resolution as compared to the average value during the science exposure. Both the between-peak event rate and the energy resolution are significantly improved compared to the previous result in Ref.~\cite{Romani}. For this calibration series, the mean photon number ($\lambda$) was 1.0 to increase statistics near zero in the short acquisition, while the science exposure used $\lambda\approx2$ to cover the full energy range of interest. The model curve is a maximum likelihood fit of photon distribution and charge transport  parameters, with results described in the text and Ref.~\cite{Romani}.}
\label{fig:laser}
\end{figure}

The calibrated detector was characterized by varying the crystal bias voltage and laser intensity while triggering on the laser coincidence signal. Figure~\ref{fig:laser} shows the reduced fill-in between laser peaks as compared to the previous result in Ref.~\cite{Romani} due to the reduced SGIR. There is still a population of fill-in events, which is well fit by an impact-ionization model with 3\% ionization probability across the 4~mm crystal thickness~\cite{phipps}. As with Ref.~\cite{Romani}, the bias scans showed linear signal scaling and constant power noise with increasing voltage (demonstrating ideal NTL amplification~\cite{Irwin2005,Luke1988,Neganov1985}).

\section{Charge Leakage}

\begin{figure}[t]
\centering
\includegraphics[width=3.25in]{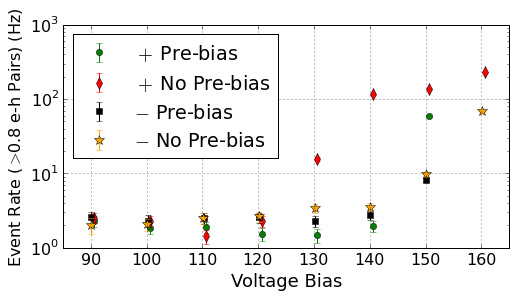}
\includegraphics[width=3.25in]{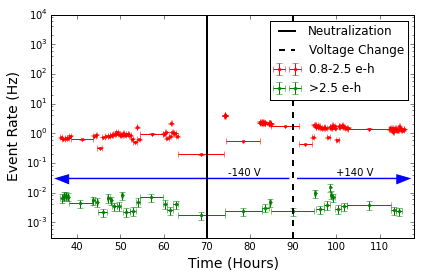}
\caption{Top: Event rate as a function of bias before and after pre-bias. Bottom: Event rate during the science exposure as a function of time. Neutralization~\cite{Romani} was performed at hour 70 (solid line) and the polarity was reversed at hour 90 (dashed line). Data points represent blocks with a fixed number of events to ensure uniform vertical error bars, with large horizontal error bars corresponding to runs separated by gaps in data taking.}\label{fig:leakage}
\end{figure}


Large electric fields used for signal amplification can auto-ionize impurities within the crystal and cause charge carriers to tunnel into the crystal at the surface, which along with SGIR, produce background events within the region of interest for DM searches. Consequently, we carefully studied the total charge leakage rate as a function of bias voltage. In these diagnostic studies, the acquisition system was configured to trigger on the laser coincidence signal, with the laser pulsed at 100\,Hz and $\lambda\approx2$. The Si crystal bias was varied in a staggered manner, increasing by 20\,V, then decreasing by 10\,V. Data were acquired at both the increasing and decreasing steps after allowing the detector to stabilize for 1 minute. This staggering enabled the study of a 10\,V pre-bias on the charge leakage of the detector. The energy spectrum of the charge leakage was determined by scanning the first half of each trace for pulses using the optimal filter. The resulting charge leakage spectrum is thus independent of the physical trigger threshold.

The measured event rate above 0.8~\eh~pairs as a function of crystal bias, largely dominated by {\em non-quantized} SGIR at lower voltages, is shown in Fig.~\ref{fig:leakage}. The event rate was $\sim$2 Hz up to $\pm$140~V ($\pm$120~V) for pre-biased (non-pre-biased) data. This event rate is 10$\times$ smaller than achieved previously, demonstrating the efficacy of our SGIR mitigations. 
Above this voltage, the {\em quantized} leakage rate increased, indicative of increased surface tunneling at the electrodes (as opposed to auto-ionization in the bulk). Full breakdown occurred around 180~V, corresponding to a field strength of $\sim$450~V/cm in the crystal bulk and in excess of $\sim$1~kV/cm near the electrode plane.

For the science exposure, the detector was pre-biased to $-$160\,V for five minutes and then biased to $-$140\,V for a minute prior to data collection to allow the detector to settle. The pre-bias was performed after each data series was acquired to ensure low charge leakage throughout the acquisition. 
As shown in Fig.~\ref{fig:leakage}, the event rate varied between 0.2--3 Hz above 0.8 \eh\ pairs. 

\section{Data Selection}

From the initial 27.4 hours of raw exposure at a detector bias voltage of $-$140\,V, a science exposure of 16.1 hours was selected based on detector performance and consistent background event rate. 
Live time and trigger efficiency were computed using the laser repetition rate and the total expected number of laser events based on the Poisson distribution of the observed laser peaks. The time associated with the observed laser events was deducted from the live time. This method allowed us to account for time variation in the energy-dependent trigger efficiency due to changes in noise environment. We verified that this method was consistent with live-time calculations using time stamps from calibration data. An exposure of 12.6 hours passed the initial, trigger- and leakage-burst cuts, yielding a science exposure of 0.49\,\gd\ for the 0.93\,g detector. 

\begin{figure}
\includegraphics[width=3.25in]{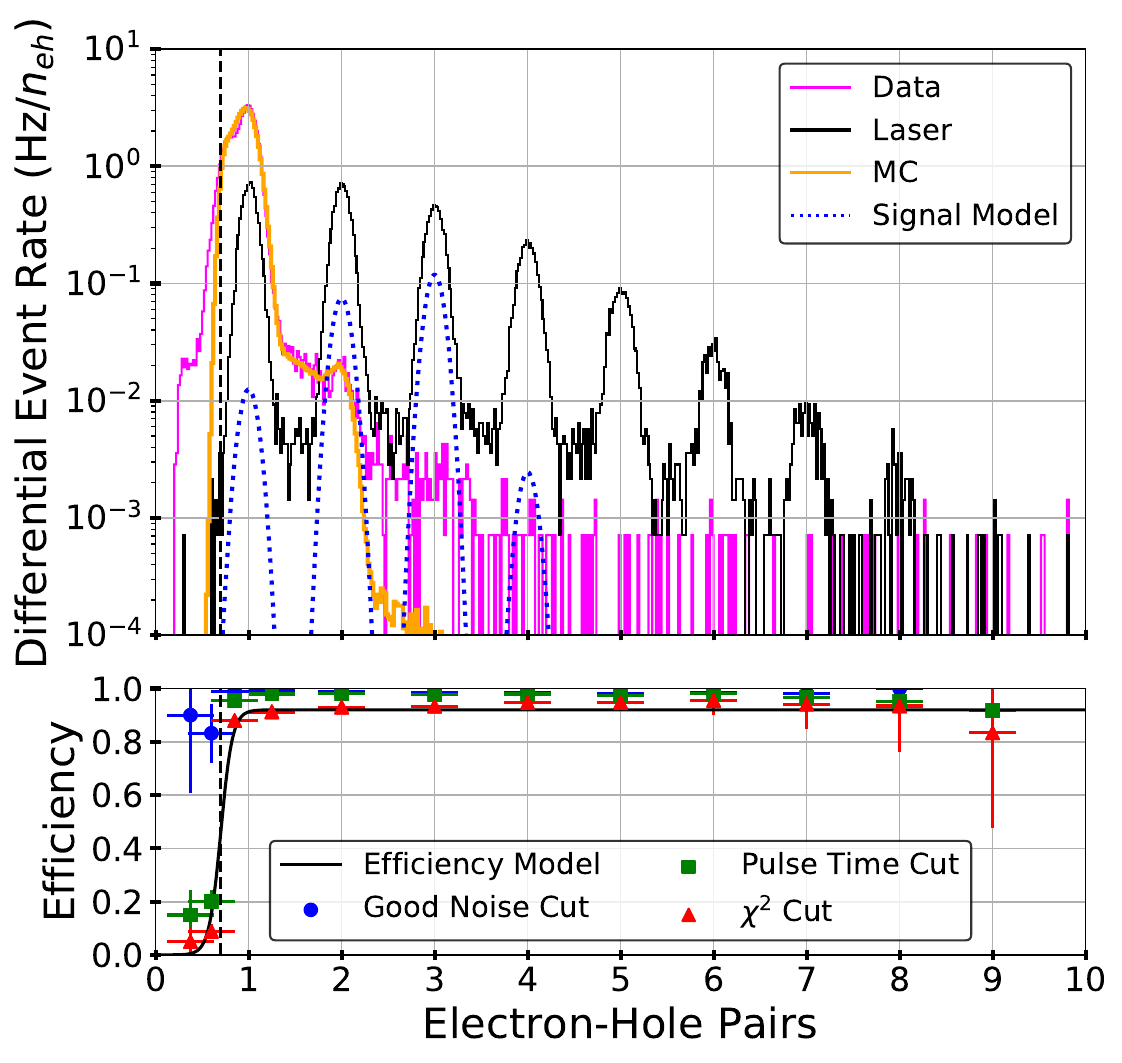}
\caption{Top: Event rate for calibration (black) and science exposure (magenta) with live time and quality cuts applied. Also shown are an impact ionization background Monte Carlo model (orange), and the signal distribution for an excluded dark photon model (dotted line) assuming $m_{V} =9.4$~eV and $\varepsilon_{\mbox{\it eff}}=5\cdot 10^{-13}$ ($\varepsilon\approx2\cdot\varepsilon_{\mbox{\it eff}}$ at 9.4 eV); the ERDM signals excluded have a similar form. Bottom: Measured cut efficiency as a function of number of \eh\ pairs along with the efficiency model used in sensitivity estimates. The dashed line in both plots shows the 50\% analysis efficiency at 0.7 \eh\ pairs.}\label{fig:data}
\end{figure}

The cut efficiency for the live time and goodness of fit cut (a basic $\chi^2$ test) as a function of the number of \eh\ pairs, $n_{eh}$, can be seen in Fig.~\ref{fig:data}, along with the laser and background spectra obtained after application of the quality and live time cuts. All of our cuts were designed to have very high efficiency, and only remove events inconsistent with the detector response, and as such are conservative. A simple background model of bulk and surface charge leakage with impact ionization, shown in Fig.~\ref{fig:data}, is an excellent fit to the data below 2 \eh\ pairs. More complex background models are expected to be capable of fitting the events above 2 \eh\ pairs.


\section{Constraints on New Physics}

We used the final 0.49 \gd\ of exposure coupled with the cut-efficiency model in Fig.~\ref{fig:data} to set limits on dark photons and ERDM. The dark photon signal model assumes kinetic mixing between the dark photon and the SM photon. The subsequent interaction of the SM photon with the material was computed according to tabulated photoelectric cross sections, giving the approximate event rate~\cite{Hochberg}
\begin{equation}
R = V_{det}\frac{\rho_{DM}}{m_{V}}\varepsilon_{\mbox{\it eff}}^2(m_V,\tilde{\sigma})\sigma_1(m_{V}), 
\end{equation}
where $V_{det}$ is the detector volume, $\rho_{DM}/m_{V}$ is the number density of DM (for this paper we assume $\rho_{DM}\sim0.3$~$\mathrm{GeV c^{-2} cm^{-3}}$~\cite{Read}), $m_{V}$ is the dark photon mass, $\varepsilon_{\mbox{\it eff}}$ is the effective kinetic mixing angle, $\tilde{\sigma}$ is the complex conductivity, and $\sigma_1(m_{V})=\mathrm{Re}(\tilde{\sigma}(m_{V}))$ is computed from the photoelectric cross section $\sigma_{p.e.}$. The kinetic mixing parameter $\varepsilon$ follows from $\varepsilon_{\mbox{\it eff}}$ after in-medium corrections as described in Ref.~\cite{Hochberg}, from which we also adopted the nominal photoelectric cross sections~\footnote{See Supplemental Material at \url{https://journals.aps.org/prl/supplemental/10.1103/PhysRevLett.121.051301}, we performed a meta-analysis of measurements of the photoelectric cross-section from the literature. This supplement includes Refs~\cite{HENKE1993181,Edwards1997547,Green,PhysRev.111.1245,Dash,Holland,Henke88,PhysRevB.49.16283,PhysRevB.2.1918,PhysRevA.37.4978,opticalprocesses,ashcroft1976solid,PhysRev.139.A560}}. 

In order to project an absorption event of known energy into our measured signal space, we adopted an ionization production model that is consistent with experimental measurements \cite{Wolf_98JAP_QYieldSolarCells, Christensen_76JAP_QYield,WilkinsonF_83JAP_QYield} and has the following mean $n_{eh}$:
\begin{equation}
\langle n_{eh}(E_{\gamma}) \rangle =\begin{cases}
0 & E_{\gamma} < E_{gap}\\
1 & E_{gap} < E_{\gamma} < \epsilon_{eh} \\
E_\gamma/\epsilon_{eh} & \epsilon_{eh} < E_{\gamma}
\end{cases}\label{eq:ionization}
\end{equation}
where $E_{gap}=1.12$~eV and $\epsilon_{eh}=3.8$~eV~\cite{Vavilov}. The probability distributions in the first two cases are delta functions. In the third case, we  generated discrete distributions with an arbitrary Fano factor, $F$, by interpolating between binomial distributions with the same $\langle n_{eh} \rangle$, but different integer number of trials. For the sensitivities shown we use the measured high energy $F$ of 0.155~\cite{OWENS_02NIM_Fano}. We also vary the $F$ used in the ionization model from its lowest mathematically possible value to 1 to estimate our sensitivity to the unmeasured ionization distribution width at low energies. Finally, we convolved the predicted \eh\ pair spectrum with the experimental resolution of 0.1 \eh\ pairs. An example of a dark photon signal ($m_{V}=9.4$~eV, $\varepsilon_{\mbox{\it eff}}=5\cdot 10^{-13}$) with this ionization model applied is superimposed on the measured spectrum in Fig.~\ref{fig:data}.

The signal induced by ERDM was calculated according to the formalism in Ref.~\cite{Essig} in which scattering rates accounting for band structure in Si are tabulated for signal modeling. The differential scattering rate is given by the function
\begin{equation}
\frac{dR}{d\ln E_R}=V_{det}\frac{\rho_{DM}}{m_{DM}}\frac{\rho_{\text{Si}}}{2m_{\text{Si}}}\bar{\sigma}_e\alpha \frac{m_{e}^2}{\mu_{DM}^2} I_{\mathrm{crystal}}(E_e;F_{DM})
\end{equation}
where $\bar{\sigma}_e\alpha$ encodes the effective DM-SM coupling, $F_{DM}$ is the momentum transfer ($q$) dependent DM form factor, $\mu_{DM}$ is the reduced mass of the DM-electron system, and $I_{\mathrm{crystal}}$ is the scattering integral over phase space in the crystal (as defined in Ref.~\cite{Essig}). We integrated this differential spectrum with Eq.~\ref{eq:ionization} to get the expected quantized spectrum, applying the same energy resolution smearing as for the dark photon signal.

\begin{figure}[ht!]
\includegraphics[width=3.25in]{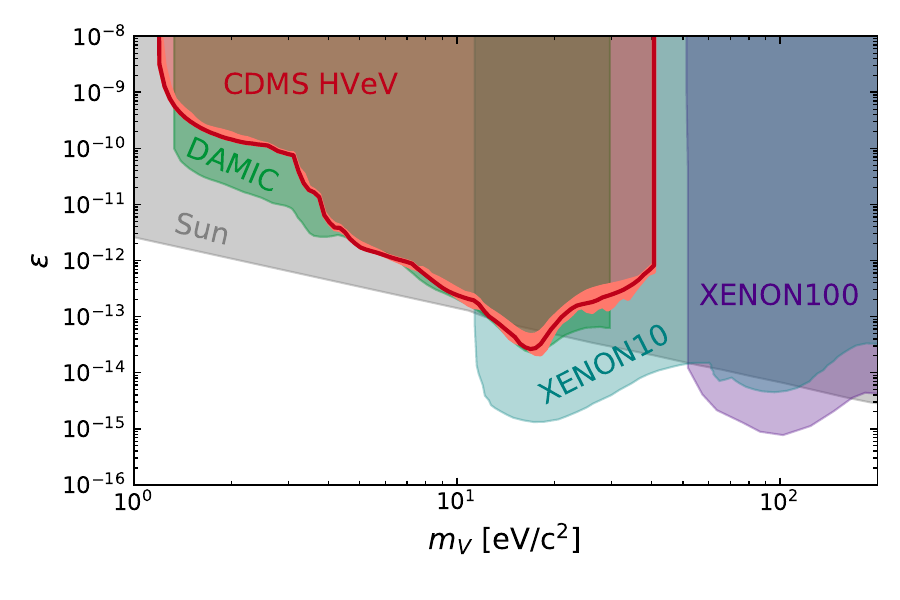}
\includegraphics[width=3.25in]{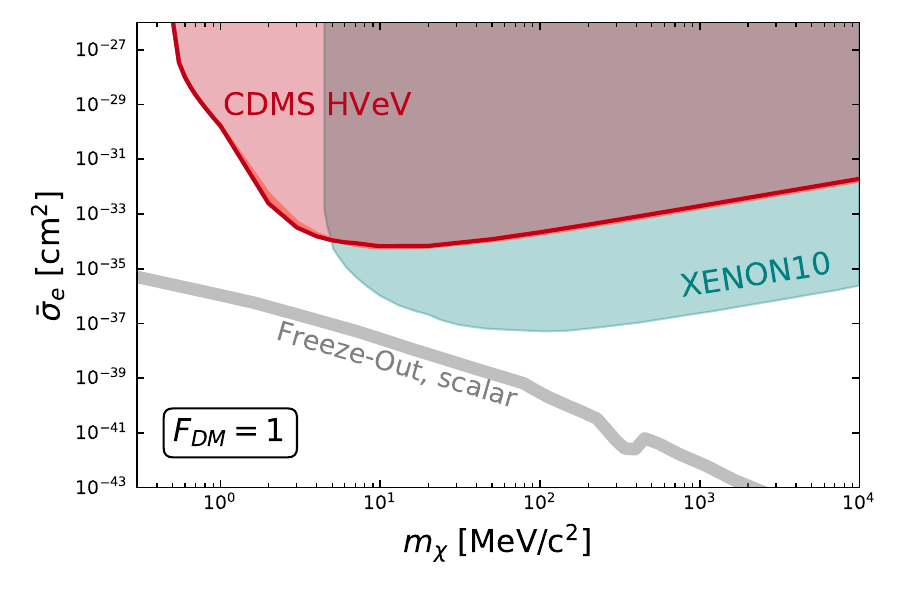}
\includegraphics[width=3.25in]{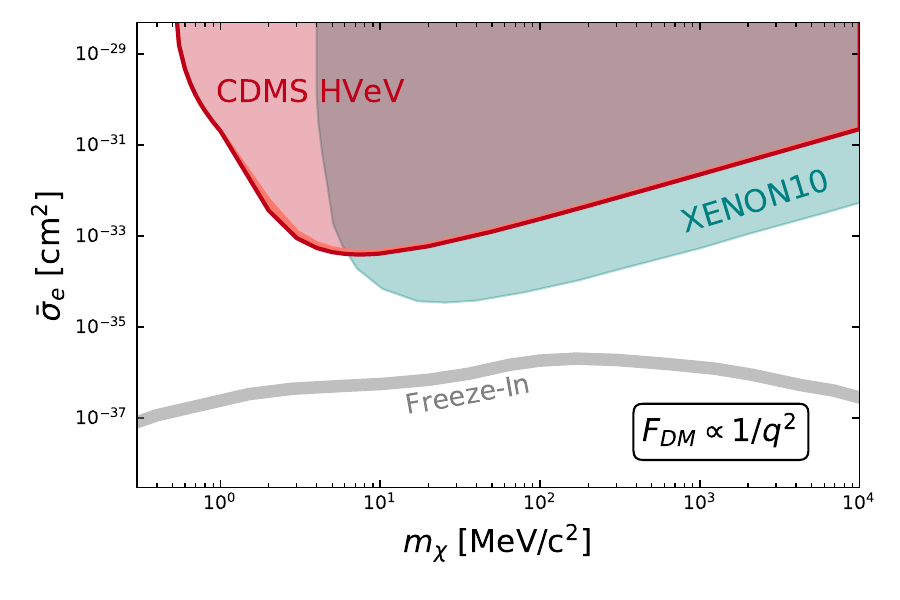}
\caption{Top: Limits on dark photon kinetic mixing compared to the results from DAMIC, XENON10 and XENON100~\cite[ and references therein]{damic}. Middle (Bottom): Limit on DM interacting with electrons via a heavy dark photon ($F_{DM}$ = 1) (ultra-light dark photon ($F_{DM} \propto 1/q^2$)) compared to the XENON10 results~\cite{Xenon10}. The red line is the limit curve with a Fano factor of 0.155. The salmon colored region indicates the systematic uncertainties due to varying the Fano factor in the ionization model between the lowest mathematically possible value and~1, as well as from uncertainties in the photoelectric cross section for dark photon absorption. For signal models as well as additional astrophysical constraints, see Ref.~\cite{CosmicVisions}.}
\label{fig:limits}
\end{figure}

We determined 90\% upper confidence limits from our data without background subtraction using the optimum interval method~\cite{Yellin,OI2007}, with the modification that we removed regions of the data $>2\sigma$ from the quantization peaks. Given that both of the DM candidates studied in this paper produced quantized signals, this ensured that the optimum interval method considered only the data likely to resemble the signals studied. Figure~\ref{fig:limits} shows the optimum interval limits for dark photon absorption and ERDM coupling via light and heavy mediators. The salmon-colored band around the exclusion limit represents the sensitivity to details of the photoelectric cross-section (below 3 eV, visible for dark photons only) and the choice of Fano factor.

\section{Discussion}

Even with this conservative analysis, DM parameter space in the mass range of 0.5--5 \MeV\, that was consistent with previously known experimental and observational bounds, has been excluded. While the XENON10 limits benefit from larger exposure above 5~\MeV, the 1.2~eV ionization energy in Si (compared to an ionization energy of 12.1~eV in Xe) allows for sensitivity to DM masses $\lesssim$~500~\keV\ for this experiment which is kinematically inaccessible to Xe targets. Furthermore, because of the minimal overburden at the experimental site (60~cm of concrete plus atmosphere), these limits are robust even for highly interacting DM candidates as long as such DM remains present in the local galactic environment \cite{Munoz_2018pzp, McDermott_PRD11_millichargedDM}. Models such as these have been hypothesized to explain recent astronomical observations~\cite{barkana}, and thus these surface-facility direct detection limits may augment other astrophysical constraints once DM survival probabilities and atmospheric absorption are more fully quantified~\cite{Berlin:2018sjs,Fraser:2018acy}. Recent results using smaller exposures in Si CCDs, with sensitivity in this same mass range, explore these surface limits further\cite{SENSEI}.

A subsequent analysis program with these data has already begun with optimized event pileup estimators and a likelihood analysis modeling of known background sources. In particular, because a large number of leakage events are non-quantized and consistent with the auto-ionization or SGIR excitation of overcharged impurities within the volume of the detector, the information between the spectral peaks can be used to constrain a physical leakage model. In addition, fill-in between the peaks at higher excitation number in both the laser calibration and background data indicates that drifting excitations have small non-negligible probabilities to be trapped-on or impact-ionize impurities. Such processes can be well-modeled with laser calibration data and by non-integer sidebands within the DM search data. This more detailed analysis is expected to produce significantly stronger DM search sensitivity than shown here. Additional calibrations, such as further studies of position dependence in the detector, will add to our understanding of the detector response and improve these background models.

Further into the future, the operation of larger-volume detectors in an underground environment as planned in the SuperCDMS SNOLAB experiment~\cite{Agnese_PRD17_SuperCDMSSensitivity} should both substantially boost the exposure of such experimental searches and decrease the leakage rate.  In particular, the new CDMS HV detectors planned for SuperCDMS SNOLAB will achieve the same NTL amplification with 8$\times$ smaller E-fields~\cite{kurinsky,Agnese_PRD17_SuperCDMSSensitivity}. Because we expect impact-ionization and surface-leakage processes to depend strongly on the E-field magnitude \cite{phipps}, we expect the primary backgrounds in the $n_{eh}\geq 2$ signal region to be substantially decreased. These types of improvements and reductions on the SGIR should decrease leakage rates further, expanding the low-mass reach of future searches. 

\section{Acknowledgements}

\begin{acknowledgements}
We thank Rouven Essig and Tien-Tien Yu for fruitful conversations and for help with understanding and using QEDark, as well as Yonit Hochberg for help understanding dark photon calculations. We also thank Gordan Krnjaic and Kathryn Zurek for fruitful theoretical discussions. Funding and support were received from the National Science Foundation, the U.S. Department of Energy, NSERC Canada, the Canada First Research Excellence Fund, MultiDark, and Michael M. Garland. This document was prepared by the SuperCDMS collaboration using the resources of the Fermi National Accelerator Laboratory (Fermilab), a U.S. Department of Energy, Office of Science, HEP User Facility. Fermilab is managed by Fermi Research Alliance, LLC (FRA), acting under Contract No. DE-AC02-07CH11359. Pacific Northwest National Laboratory is operated by Battelle Memorial Institute under Contract No. DE-AC05-76RL01830 for the U.S. Department of Energy. SLAC is operated under Contract No. DEAC02-76SF00515 with the U.S. Department of Energy.
\end{acknowledgements}

\bibliographystyle{apsrev4-1-JHEPfix-autoEtAl}
\bibliography{refsEH}

\clearpage

\onecolumngrid
\begin{center}
{\large \bf Erratum: First Dark Matter Constraints from a SuperCDMS Single-Charge Sensitive Detector} \\[0.25cm]
See below for author list
\end{center}
\vspace{0.5cm}
\twocolumngrid

\begin{figure}
\centering
\includegraphics[width=3.1in,clip=True,trim=20 20 0 10]{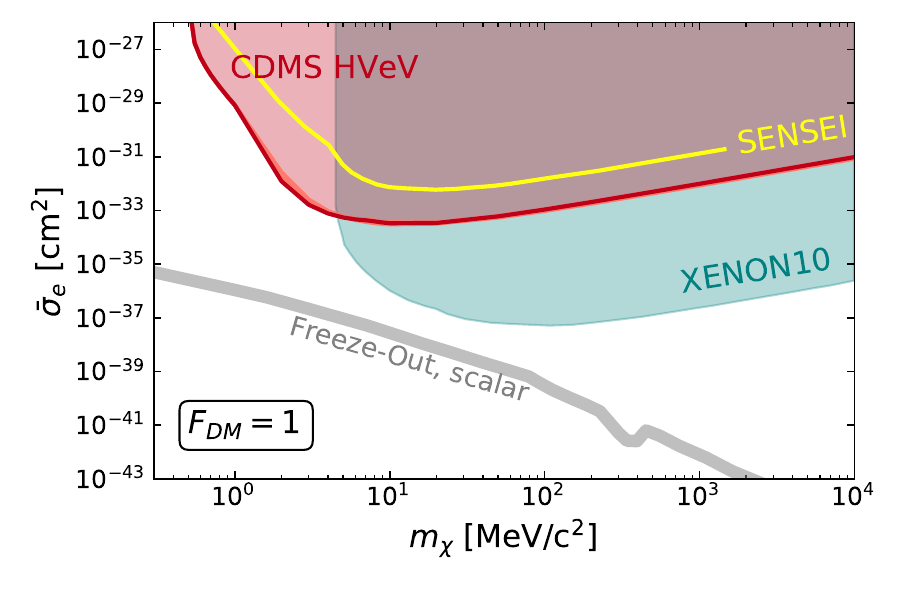}
\includegraphics[width=3.1in,clip=True,trim=20 20 0 10]{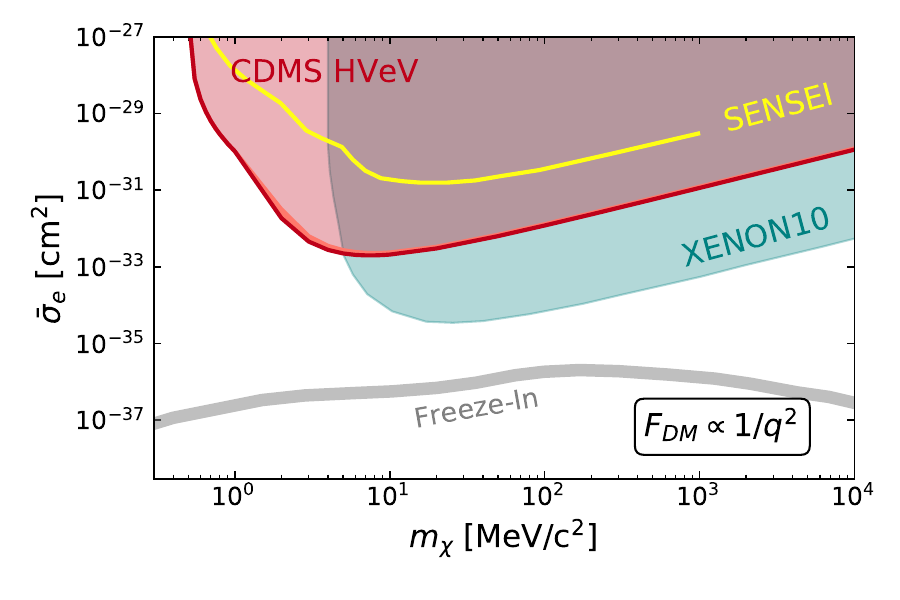}
\caption{Corrected limit on DM particle interacting with electrons via a heavy dark photon (Top, $F_{DM}$ = 1) or an ultra-light dark photon (Bottom, $F_{DM} \propto 1/q^2$)) compared to the XENON10 and SENSEI results~\cite{Xenon10,SENSEI}. The red line is the limit curve with a Fano factor of 0.155 in the ionization model. The salmon colored region indicates the systematic uncertainties due to varying the Fano factor between the lowest mathematically possible value and 1. For signal models as well as additional astrophysical constraints, see \cite{CosmicVisions}. All limits as shown here assume a local DM density of 0.3\,GeV/cm$^3$.}
\label{fig:limitsCorr}
\end{figure}

\begin{figure}
\centering
\includegraphics[width=3.1in,clip=True,trim=20 20 0 0]{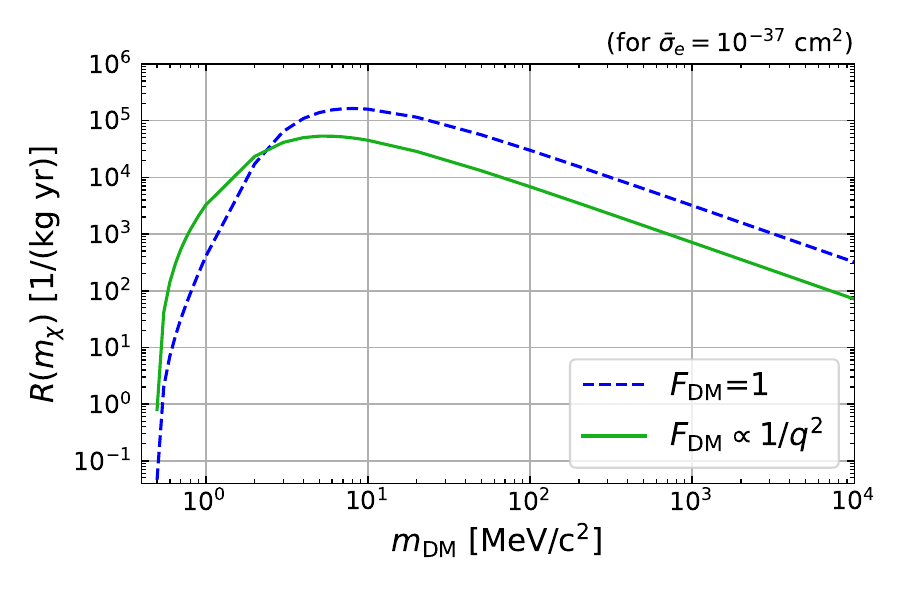}
\caption{Total rates $R(m_{\chi})$ of DM--electron scattering in silicon for two DM form factors, $F_{DM}$, corresponding to different DM models. The blue dashed (green solid) line assumes a heavy (ultra-light) dark photon mediator. The rates are the yearly average for a local DM density of 0.3\,GeV/cm$^3$ and are calculated with QEdark \cite{qedark}.}
\label{fig:rate}
\end{figure}

In our publication describing the search for Dark Matter (DM) using a cryogenic Si chip sensitive to single electron-hole pairs \cite{HVeV}, the differential scattering rate of DM particles with electrons for a given DM particle mass $m_{\chi}$, $dN/dE(m_{\chi})$, was computed using the output of the publicly available QEdark notebook \cite{qedark}. From this notebook we obtained an array in 0.2 eV bins of $\Delta N_i$, the expected number of events in bin $i$, for a fixed dark matter density, data acquisition time, detector mass, and cross section. In our linear approximation $\Delta N_i $ is related to $dN/dE$ as $\Delta N_i \approx dN/dE \times \Delta E_i = dN/dE \times 0.2$, with the recoil energy, $E$, taken to have units of eV; so $dN/dE \approx 5\Delta N_i$. Due to a miscommunication, this factor of 5 binning correction was applied twice, yielding a $dN/dE$ that was five times higher than it should have been and an upper limit cross section that was five times too strong. 

An updated version of Fig.~4 middle and bottom of the original publication is provided in this Erratum as Fig.~\ref{fig:limitsCorr}. These new figures also include the limits observed by the SENSEI Collaboration which were published simultaneously with our original publication in \cite{SENSEI}. For ease of future comparison, Fig.~\ref{fig:rate} has been added to this Erratum showing the total DM--electron scattering rate in silicon at $\bar{\sigma}_e=10^{-37}$\,cm$^2$ for the probed DM models. This rate was calculated with the QEdark notebook downloaded on October 28, 2018 and forms the basis of the results shown in Fig.~\ref{fig:limitsCorr}. The notebook has not changed since \cite{HVeV}.

\vspace{0.5cm}

\begin{acknowledgements}
We thank Rouven Essig for useful discussions.
\end{acknowledgements}

\clearpage

\end{document}